# Virtual Web Based Personalized PageRank Updating

Bo Song
*Department of Electrical Engineering and Computer Science*
*University of Missouri-Columbia*
Columbia, Missouri
bosong@mail.missouri.edu

Xiaobo Jiang
*Department of Electrical Engineering and Computer Science*
*University of Missouri-Columbia*
Columbia, Missouri
xjd3b@mail.missouri.edu

Xinhua Zhuang
*Department of Electrical Engineering and Computer Science*
*University of Missouri-Columbia*
Columbia, Missouri
zhuangx@missouri.edu

*Abstract*—Growing popularity of social networks demands a highly efficient Personalized PageRank (PPR) updating due to the fast-evolving web graphs of enormous size. While current researches are focusing on PPR updating under link structure modification, efficiently updating PPR when node insertion/deletion involved remains a challenge. In the previous work called *Virtual Web* (VW), a few VW architectures are designed, which results in some highly effective initializations to significantly accelerate PageRank updating under both link modification and page insertion/deletion.

In the paper, under the general scenario of link modification and node insertion/deletion we tackle the fast PPR updating problem. Specifically, we combine VW with the *TrackingPPR* method to generate initials, which are then used by the *Gauss-Southwell* method for fast PPR updating. The algorithm is named *VWPPR* method. In extensive experiments, three real-world datasets are used that contain 1~5.6M nodes and 6.7M~129M links, while a node perturbation of 40k and link perturbation of 1% are applied. Comparing to the more recent *LazyForwardUpdate* method, which handles the general PPR updating problem, the *VWPPR* method is 3~6 times faster in terms of running time, or 4.4~10 times faster in terms of iteration numbers.

## I. INTRODUCTION

Personalized PageRank (PPR) is a variation of the renowned Google's PageRank algorithm [22]. It measures the importance of the nodes based on the link connectivity information of the network, while taking the users' preferences into consideration. It was first proposed to provide personalized search results for general web search [13]. As the social networks thrives, the PPR has widespread use in social searches, especially in the recommendations systems such as friend recommendations as in [12], or video recommendations [6]. As the real-world networks are often enormous and evolving rapidly, updating PPR by re-compute from scratch is impractical due to the expensiveness of the computation. An efficient PPR updating algorithm which can effectively utilize the previous results to accelerate the computation over the updated networks is therefore demanded. Previous works on computing PPR for evolving networks are mainly focused on the link structure evolving problem, that the nodes are often assumed static [4][5][18][19][21][26]. However, user actions that cause node changes are not rare: addition/removal of the web pages, or creation/deletion of the social network accounts happen momentarily. In this work, we proposed an efficient PPR updating algorithm aiming to solve the general case when both link structure and node evolvements are involved.

Previous works on PPR computing can be mainly categorized into three kinds: the power method [13][22], the Monte-Carlo method [3] and the Local-push method [1][7]. The power method is the base-line method for PageRank/PPR computing, but the $\Omega(|E|)$ complexity makes it impractical to be used for PPR updating, especially for such massive amount of users. The Monte-Carlo method is a random-walk-simulation-based method, that the PPR score of a target node $t$ for a given source node s is approximated by the frequency of $t$ being visited during some $k$ random-walk simulations initiated from s. In [4], Bahmani et al. use Monte-Carlo method to track PPR on an evolving network. The approach maintains multiple random-walk segments, from which the PPR scores can be quickly approximated. As the link structure modifies, the segments are reconstructed. The method requires massive pre-processing and storage for random-walk segments. In [18], Lofgren et al. proposed pair-wise approach by Monte-Carlo method, which trace both forwardly from the source node and backwardly from the target node with multiple random-walk simulations to estimate the PPR. This method represents a novel approach for fast PPR computation in social networks, but the demand for pre-processing and storage remains high. In their work based on undirected graph [19], the Local-push method replaced the forward trace to save the pre-computations and storages needed in [18]. The convergence rate and accuracy are two major issues of the Monte-Carlo method as mentioned by Avrachenkov et al. in [3] and is testified experimentally by Ohsaka et al. in [21]. bring people's attention for PPR updating recently. Recently, the researchers developed efficient PPR algorithms based on the Local-push method [21][19][26], for updating networks with minor link-structure perturbations. In local-push method, small amendments are made to the previous results according to the link perturbations, so the update computing is localized to the perturbation and the real-time PPR updating becomes feasible. In [21], Ohsaka et al. proposed the *TrackingPPR* algorithm which utilizes the differences between the transition matrices of the previous and the updated web, to form a revised residual vector as the initialization for the next local-push PPR computing, for real-time PPR tracking purpose. Zhang et al. proposed *LazyForwardUpdate* algorithm in [26] for undirected graphs. The PPR is updated per each single link modification. Given a link inserted or removed, the previous PPR and residual vectors are amended accordingly, which are then used as initials for the next PPR update computing. The experiment result shows that the *LazyForwardUpdate* method is 2-3 times faster than *TrackingPPR* while both of them achieved similar accuracy. We are interested in their solution for the node insertion/deletion problem. In their work, updating for a node insertion can be done by inserting the node's in/out-links into the network one-by-one, and then invoke the proposed algorithm to update each of the link insertions, or reverse for the node deletions.

In the paper, we use *Virtual Web* (VW) [25] to tackle the PPR updating problem when link and node insertion/deletion occur simultaneously during web evolving. Specifically, we combine the idea of VW with the *TrackingPPR* method [21] to generate an initial PPR estimation and correspondingly an initial residual vector from the previous PPR and residual vector. The initials are then used in the Gauss-Southwell method (a variation of the local-push method used in [21]) for fast computing of the current PPR. The proposed algorithm is named *VWPPR* method. Extensive experiments are conducted and compared to the *LazyForwardUpdate* method. The latter is the only recent work to our best knowledge that proposed a solution to handle the simultaneous link modification and node insertion/deletion. Experiments are conducted on three real-world datasets with size of 1M ~ 5.6M nodes and 6.7M ~ 129M edges. In the meantime, node insertion/deletion each of 20k and link insertion/deletion together up to 1% of the total edges are applied. The results show that with nearly the same accuracy, the proposed *VWPPR* method is 3~6 times faster than the *LazyForwardUpdate* method in terms of running time, or 4.4~10 times faster in terms of iteration number.

The rest of the paper is organized as follows. In Section 2, we provide some theoretical fundamentals and analysis of the PPR, as well as the Gauss-Southwell method for PPR computing. It is mathematically showed that, given a transition matrix, it is the initialization to decide the convergence rate. The proposed algorithm is introduced in Section 3, in which we demonstrated the theories we used to guide the design of the algorithm. Also, the general web evolving model is explained. In Section 4, experiments are conducted to compare the proposed algorithm with the cutting-edge method for PPR updating. Section 5 concludes the paper.

## II. PRELIMINARIES

### A. Definition of Personalized PageRank (PPR)

The PPR differs from the original PageRank algorithm [18] by the preference vector, which is a uniform probability vector in regular PageRank but an arbitrary stochastic vector in PPR which biased to the user's preference. In this work, we follow the egocentric definition of the PPR despite of [18], which has widespread usage in social networks [18][19][26]. Given a network $G = (V, E)$ where $V$ is the set of nodes and $E$ is the set of edges (or links). For a source node $s$ in $V$, its associated PPR vector $\pi_s$ is defined by:

$$\pi_s = \alpha \pi_s P + (1-\alpha)\mu_s \quad (1)$$

where $\mu_s = e_s$ is an indicator vector with all elements being zero except the $s$-th element being 1. $\alpha$ is the random-jump constant that is conventionally set to 0.85. The transition matrix $P$ of graph $G$ is defined as:

$$P: \begin{cases} p_{ij} = 1/d_i, & \forall (i \to j) \in E \\ p_{ij} = 0, & \text{otherwise} \end{cases} \quad (2)$$

where $d_i$ is the out-degree of node $i$, i.e., the total number of links originated from $i$. It's easy to see that (1) is equivalent to (4) as below:

$$(1-\alpha)\mu_s = \pi_s (I - \alpha P) \quad (3)$$

$$\pi_s = (1-\alpha)\mu_s (I - \alpha P)^{-1}$$

$$= \lim_{t \to \infty}(1-\alpha)\mu_s \sum_{0 \le k < t}(\alpha P)^k \quad (4)$$

where the matrix $(I - \alpha P)$ is invertible since $\|\alpha P\|_1 \le \alpha < 1$. As known, the $l_1$ norm of a vector is defined by the sum of the absolute values of its elements, while the $l_1$ norm of a matrix by the maximum of the $l_1$ norms of its row vectors. Moreover, the $l_1$ norm of a vector multiplied by a matrix *from right* is no larger than the product of the $l_1$ norm of the vector and the $l_1$ norm of the matrix.

### B. Personalized PageRank vs. PageRank

The regular PageRank $\pi$ over $P$ is defined by:

$$\pi = \alpha \pi P + (1-\alpha)\mu \quad (5)$$

where $\mu$ is the uniform probability vector, i.e., $\mu = [1/n]_{1 \times n}$. Interestingly, the regular PageRank (PR) equals the average of the egocentric PPR:

**Lemma 1**. Given a directed graph $G = (V, E)$ with $n = |V|$ nodes and $m = |E|$ edges. Its regular PageRank is the average of the egocentric Personalized PageRank of *n* nodes.

**Proof**. From (1) we may derive

$$\frac{1}{n}\sum_{\forall s \in V} \pi_s (I - \alpha P) = \frac{1}{n}\sum_{\forall s \in V}(1-\alpha)\mu_s = (1-\alpha)\mu \quad (6)$$

$$\frac{1}{n}\sum_{\forall s \in V} \pi_s (I - \alpha P) = \pi(I - \alpha P) \Rightarrow \pi = \frac{1}{n}\sum_{\forall s \in V} \pi_s$$

Thus Lemma 1 is proved.

### C. Gauss-Southwell Method of PPR

The *Gauss-Southwell* (GS) method [21], or *Local-push* method, is an iterative algorithm for PR/PPR computing. Recent works using GS or Local-push method for PPR calculation can be found in [21][19][26]. GS method is an alternative to the Power-Iteration method [22] for PR/PPR computing, which transforms the convergence from checking the $l_1$ -distance of the PPR vectors of two consecutive iterations, to checking the maximum-norm of a residual vector $r$. Briefly, the iterative process on the residual vector of the GS method can be described by the following equation:

$$r_s^{(t)} = (1-\alpha)\mu_s - \pi_s^{(t)}(I - \alpha P) \quad (7)$$

where $\pi_s^{(t)}$ denote the PPR $\pi_s$ at $t$-th iteration and $r_s^{(t)}$ denote its corresponding residual vector. The initialization $\pi_s^{(0)} = 0$ by default, thus $r_s^{(0)} = (1-\alpha)e_s$. The detailed algorithm is presented in Algorithm 1.

**Algorithm 1.** *GaussSouthwellPPR*$(P, \pi_s^{(0)}, r_s^{(0)}, \epsilon)$

**Inputs**: transition matrix $P$, error-bound $\epsilon$
**Initialization**: $\pi_s^{(0)} = 0, r_s^{(0)} = (1-\alpha)\mu_s$
1:    $t \leftarrow 0$
2:    **while** $\exists i \in V$ so that $|r_s^{(t)}[i]| > \epsilon$
3:       $\pi_s^{(t+1)} = \pi_s^{(t)} + r_s^{(t)}[i]e_i$
4:       $r_s^{(t+1)} = r_s^{(t)} - r_s^{(t)}[i]e_i + \alpha r_s^{(t)}[i]e_i P$
5:       $t$++;
6:    **end while**
7:    **return** $\langle \pi_s^{(t)}, r_s^{(t)} \rangle$

*$r_s^{(t)}[i]$ is the $i$-th element in $r_s^{(t)}$.

### D. Source-Node-Originated Path and Reachable Nodes

A path $\mathbb{P}(s \sim j)$ from source node $s$ to $j$ exists if $j$ is reachable from $s$ following some available links. Given the source node $s$, $V$ can be thus divided into two subsets: $\mathcal{S}$ (for nodes reachable from $s$) and $\mathcal{U}$ (for nodes unreachable from $s$), $\mathcal{S} \cup \mathcal{U} = V$ and $\mathcal{S} \cap \mathcal{U} = \emptyset$, so that $\mathbb{P}(s \sim v)$ exists for any $v \in \mathcal{S}$ and $\mathbb{P}(s \sim u)$ does not exist for any $u \in \mathcal{U}$. Let the transition matrix $P$ be re-arranged so that the top rows (and columns) are for $\mathcal{S}$-nodes and the rest for $\mathcal{U}$-nodes. It's clear that each node in $\mathcal{S}$ is reachable from source node $s$ while each node in $\mathcal{U}$ is not linked from any node in $\mathcal{S}$. Let

$$P = \begin{bmatrix} P_1 & P_2 \\ P_3 & P_4 \end{bmatrix} \quad (8)$$

where $P_1$ is the submatrix corresponding to the reachable nodes $\mathcal{S}$ and $P_2 = 0$. We then have the following lemma:

**Lemma 2.** Let the PPR vector of $\langle G, s \rangle$ be $\pi_s = (\pi_s^\mathcal{S}, \pi_s^\mathcal{U})$, where $\pi_s^\mathcal{S}$ and $\pi_s^\mathcal{U}$ are the sub-vectors in correspondence to $\mathcal{S}$ or $\mathcal{U}$, respectively. Then, $\pi_s^\mathcal{U} = 0$.

*Proof.* By PPR definition in (1), we may easily arrive at

$$\pi_s^\mathcal{S} = \alpha(\pi_s^\mathcal{S} P_1 + \pi_s^\mathcal{U} P_3) + (1-\alpha)\mu_s^\mathcal{S}$$
$$\pi_s^\mathcal{U} = \alpha\pi_s^\mathcal{U} P_4 + (1-\alpha)\mu_s^\mathcal{U}$$

Since $\mu_s^\mathcal{U}$ is a zero vector, and $(I - \alpha P_4)$ is invertible, we have $\pi_s^\mathcal{U} = 0$. Finally, $\pi_s^\mathcal{S} = \alpha\pi_s^\mathcal{S} P_1 + (1-\alpha)\mu_s^\mathcal{S}$, i.e., $\pi_s^\mathcal{S}$ is the PPR of $\langle G_S, s \rangle$, where $G_S$ is the graph of $\mathcal{S}$.

Lemma 2 demonstrates that PPR of $P$ can be simply given by PPR of $P_1$ over a reduced subgraph $\mathcal{S}$ plus $k$ zeros, where $k$ is the size of the unreachable nodes from $s$. In general, the cost of forming $\mathcal{S}$ is not ignorable. However, the GS method will never run over $\mathcal{U}$. As seen, GS algorithm can only start from the source node $s$ because $r_s^{(0)} = (1-\alpha)e_s$. Assume at time $t-1$ a node $i \in \mathcal{S}$ is picked by the algorithm. Line 3 in Algorithm 1 means only the $i$-th element of $\pi_s^{(t-1)}$ will be changed. Line 4 indicates that only components in $r_s^{(t-1)}$ corresponding to the out-neighbors of $i$ will be changed. Since there exists no path from $s$ to $\forall u \in \mathcal{U}$, there is no chance to ever visit the subset $\mathcal{U}$. This clearly is an advantage of the GS method for single-sourced PPR computing.

### E. Analysis of the Gauss-Southwell Method

In this section, we will theoretically analyze the GS method for Personalized PageRank updating.

**Lemma 3.** $r_s^{(t)} = (\pi_s - \pi_s^{(t)})(I - \alpha P)$ holds for GS method (before its convergence).

*Proof.* By GS method, we have

$$r_s^{(0)} = (1-\alpha)\mu_s - \pi_s^{(0)}(I - \alpha P)$$
$$= \pi_s(I - \alpha P) - \pi_s^{(0)}(I - \alpha P) = (\pi_s - \pi_s^{(0)})(I - \alpha P)$$
$$r_s^{(t+1)} = r_s^{(t)} - r_s^{(t)}[i]e_i(I - \alpha P)$$
$$= r_s^{(t)} - (\pi_s^{(t+1)} - \pi_s^{(t)})(I - \alpha P)$$

Suppose at iteration $t$, $r_s^{(t)} = (\pi_s - \pi_s^{(t)})(I - \alpha P)$ ($t > 0$). Then,

$$r_s^{(t+1)} = (\pi_s - \pi_s^{(t)})(I - \alpha P) - (\pi_s^{(t+1)} - \pi_s^{(t)})(I - \alpha P)$$
$$= (\pi_s - \pi_s^{(t+1)})(I - \alpha P)$$

By induction, $r_s^{(t)} = (\pi_s - \pi_s^{(t)})(I - \alpha P)$ holds.

**Lemma 4.** $\|\pi_s - \pi_s^{(t)}\|_1 \leq \|r_s^{(t)}\|_1 \frac{1}{(1-\alpha)}$ holds for GS method (before its convergence).

*Proof.* By Lemma 3, $\pi_s - \pi_s^{(t)} = r_s^{(t)}(I - \alpha P)^{-1}$. Thus,

$$\|\pi_s - \pi_s^{(t)}\|_1 \leq \|r_s^{(t)}\|_1 \|(I - \alpha P)^{-1}\|_1 \leq \|r_s^{(t)}\|_1 \frac{1}{(1-\alpha)}$$

As showed in [21], the following holds when GS algorithm converges at iteration $\tau$:

$$\|r_s^{(\tau)}\|_1 \leq \|r_s^{(0)}\|_1 - (1-\alpha)\sum_{t=0}^{\tau-1}|r_s^{(t)}[i_t]|$$
$$\leq \|r_s^{(0)}\|_1 - (1-\alpha)\epsilon\tau$$

The equation above shows that a good initial $\pi_s^{(0)}$, which produces a smaller residual $\|r_s^{(0)}\|$, would effectively reduce the iteration number $\tau$ needed till convergence, to achieve faster computing speed. Let residual $r_s = (1-\alpha)\mu_s - \pi_s(I - \alpha P)$ associated with the converged previous PPR $\pi_s$ to be used to initialize the computation of the updated PPR $\pi_s^*$. That is, $\pi_s^{*(0)} = \pi_s$. The initial residual for $\pi_s^{*(0)}$ can be set as:

$$r_s^{*(0)} = (1-\alpha)\mu_s - \pi_s(I - \alpha P^*)$$
$$= r_s + \alpha\pi_s(P^* - P) \quad (9)$$

Supposedly, only a small number of rows are different from $P$ to $P^*$ (if we recompute PPR after only some small perturbations). Thus, most of the rows in matrix $(P^* - P)$ are zeros, so the complexity of computing $\alpha\pi_s(P^* - P)$ is significantly reduced. This method is named *TrackingPPR* as described in [21].

III. VIRTUAL INITIALS FOR PPR UPDATING

In [25], we proposed a concept called *Virtual Web* to tackle the general PageRank updating as web evolution involves both link structure modification and node insertion/deletion. The concept is developed based on the theoretical analysis and empirical observation that the previous PageRank can serve as an effective initialization for PageRank updating, when only link structure modification occurs from the old web to the new web. The concept can be used for PPR updating when multiple node/link insertions and deletions occur simultaneously.

*A. Virtual Web Design and Virtual Initial for PPR*

Let the current network $W^*$ be evolved from the previous network $W$ through the insertion of $a$ new nodes and deletion of $d$ old nodes plus some link modifications. Let $P$ be the transition matrix of $n \times n$ for $W$ and $\pi_s$ be its PPR. Let $P^*$ denote the transition matrix of $n^* \times n^*$ for $W^*$ and $\pi_s^*$ be its PPR, where $n^* = n + a - d$. Without loss of generality, we may assume two transition matrices $P$ and $P^*$ being organized as follows with:

$$P = \begin{bmatrix} P_0 & P_{0,d} \\ P_{d,0} & \hat{P}_d \end{bmatrix} \tag{10}$$

$$P^* = \begin{bmatrix} P_0^* & P_{0,a}^* \\ P_{a,0}^* & \hat{P}_a^* \end{bmatrix} \tag{11}$$

where $P_0, P_0^* \in \mathbb{R}_{(n-d)\times(n-d)}$. $\hat{P}_a \in \mathbb{R}_{a \times a}$ and $\hat{P}_d \in \mathbb{R}_{d \times d}$ are the sub-matrices respect to the inserted or deleted nodes. $P_0$ and $P_0^*$ represent the same set of nodes that remained during the evolvement, which are of the same dimensional but have possibly different link structures. $P_{0,d} \in \mathbb{R}_{(n-d)\times d}$ and $P_{d,0} \in \mathbb{R}_{d \times (n-d)}$ are the sub-matrices of $P$ that connecting $P_0$ and $\hat{P}_d$; similarly, $P_{0,a}^* \in \mathbb{R}_{(n-d) \times a}$ and $P_{a,0}^* \in \mathbb{R}_{a \times (n-d)}$ are the sub-matrices of $P^*$ that connecting $P_0^*$ and $\hat{P}_a$;

For PPR updating, we define a virtual web $W_v$ by simply adding a virtual transition matrix $P_a \in \mathbb{R}_{a \times a}$ to $P$. We then define the second virtual web $W_v^*$ by simply adding a virtual transition matrix $P_d \in \mathbb{R}_{d \times d}$ to $P$. The transition matrices of those two virtual webs, $P_v$ and $P_v^*$, are as follows:

$$P_v = \begin{bmatrix} P & 0_{n \times a} \\ 0_{a \times n} & P_a \end{bmatrix} \tag{12}$$

$$P_v^* = \begin{bmatrix} P^* & 0_{n^* \times d} \\ 0_{d \times n^*} & P_d \end{bmatrix} \tag{13}$$

An intuitive example of web evolution and the formation of the virtual webs is demonstrated in Fig 1. Basically, $P_v$ and $P_v^*$ are of the same dimensional, and we assume a node perturbation of small size from $P$ to $P^*$, so that the PPR over $P_v$ can be utilized as an efficient initialization for computing PPR $\pi_s^{v*}$ over $P_v^*$. Next, the true PPR $\pi_s^*$ can be easily separated from $\pi_s^{v*}$. The construction of the virtual web for PPR computing is inspired by the decomposition in [2] and Lemma 2, detailed explanation of how it can be applied on general PageRank updating is illustrated in [25].

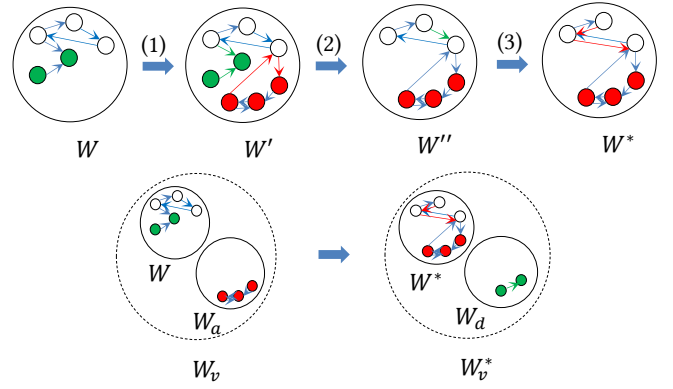

Fig. 1. The web evolvement. Up: (1) Node Insertion (red colored): some nodes together with some links are inserted into the original web W, resulting in $W'$; (2) Node Deletion (green colored): some nodes together with relevant links removed, resulting in $W''$; (3) Link structure modification (red links inserted and green link removed) occurs among the remaining nodes (white colored), resulting in $W^*$ finally. Down: the inserted nodes $W_a$ and the old web $W$ form the first virtual web $W_v$; Then, $d$ deleted nodes are removed from $W_v$, which forms the virtual web $W_v^*$. $W_v$ and $W_v^*$ share the same set of vertices, the only difference in between lies in their link structures.

Let $\pi_s^v$ denote the PPR of a given source $s$ over $P_v$, where $s$ is a node in $P_0$. Obviously, $\pi_s^v = \alpha \pi_s^v P_v + (1-\alpha)\mu_s^v$, where $\mu_s^v$ is a vector of all zero elements except the element at the source node $s$ is 1. It can be easily induced from Lemma 2 that independent of choices for $P_a$,

$$\pi_s^v = (\pi_s, 0_{1 \times a}) \tag{14}$$

The vertex order in $\pi_s^v$ can be re-arranged with no name change as follows:

$$\pi_s^v = (x, 0_{1 \times a}, x_d) \tag{15}$$

where $\pi_s = (x, x_d)$; $x \in \mathbb{R}_{1 \times (n-d)}$ is part of PPR $\pi_s$ while $x_d \in \mathbb{R}_{1 \times d}$ is part of $\pi_s$, relevant to the deleted nodes. Let the corresponding residual be $r_s = (\gamma, \gamma_d)$. Now we are ready to define a virtual PPR $x_v$ and a virtual residual $\gamma_v \in \mathbb{R}_{1 \times n^*}$, as follows:

$$x_v = (x, 0_{1 \times a}), \gamma_v = (\gamma, 0_{1 \times a}) \tag{16}$$

According to the arrangement of $\pi_s^v$ in (15), we can re-arrange the virtual transition matrix $P_v$ into following form with no name change:

$$P_v = \begin{bmatrix} P_0 & 0 & P_{(n-d) \times d} \\ 0 & P_a & 0 \\ P_{d \times (n-d)} & 0 & P_d \end{bmatrix} \tag{17}$$

And let $P^o$ be the matrix at the upper-left corner of $P_v$:

$$P^o = \begin{bmatrix} P_0 & 0 \\ 0 & P_a \end{bmatrix} \tag{18}$$

where $P^o \in \mathbb{R}_{n^* \times n^*}$ is of the same dimensional to $P^*$. Hence, there exists only link structure difference from $P^o$ to $P^*$. To apply GS algorithm to calculate PPR for the current web $W^*$, we start from:

$$\pi_s^{*(0)} = x_v \qquad (19)$$

$$r_s^{*(0)} = \gamma_v + \alpha x_v(P^* - P^o) \qquad (20)$$

Even though $P_a$ plays no role in (16), using $P_a = \hat{P}_a^*$ would further simplify (20) as below,

$$r_s^{*(0)} = (\gamma + \alpha x(P_0^* - P_0), \alpha x P_{0,a}^*) \qquad (21)$$

$r_s^{*(0)}$ and $\pi_s^{*(0)}$ are then used as initialization to compute the PPR over $P^*$. The algorithm is presented as below:

---
**Algorithm 2.** VWPPR($P, \pi_s^{*(0)}, r_s^{*(0)}, \epsilon$)

---
**Inputs**: transition matrix $P$, error-bound $\epsilon$

**Initialization**: $\pi_s^{*(0)}, r_s^{*(0)}$ are given by (19) & (21).

1: $\langle \pi_s^{*(t)}, r_s^{*(t)} \rangle \leftarrow GaussSouthwellPPR(P, \pi_s^{(0)}, r_s^{(0)}, \epsilon)$;
2: **return** $\langle \pi_s^{*(t)}, r_s^{*(t)} \rangle$

---

The complexity of the proposed algorithm follows the *TrackingPPR* method in [21].

## IV. EXPERIMENTS

### A. Dataset

*1) Datasets:* Three public available real-world social network datasets are used: the *enwiki-2018* dataset is graph of the English Wikipedia acquired by the Laboratory for Web Algorithmics [26]; and the *soc-pokec* and *soc-LiveJournal1* datasets are obtained from the Stanford Large Network Dataset Collection [24]. The details of the datasets are listed in Table I.

TABLE I.  STATISTICS OF THE DATASETS

| Datasets | $|V|$ | $|E|$ |
|---|---|---|
| soc-pokec | 1.63M | 31M |
| soc-LiveJournal1 | 4.8M | 69M |
| enwiki-2018 | 5.62M | 129M |

*2) Node Insertion/Deletion*: For each given dataset $\mathcal{S}$, we first select a random node as a seed. A subset $\mathcal{A}$ of nodes is fetched from that seed through Breadth-First-Search (BFS). The original web $\mathcal{W}^O$ before any node insertion is formed by removing $\mathcal{A}$ from $\mathcal{S}$, that $\mathcal{A}$ represents the nodes to be inserted later. Next, we remove a subset $\mathcal{D}$ of nodes that are randomly selected from $\mathcal{W}^O$, that $\mathcal{D}$ represents the nodes to be deleted during the web evolvment. The updated web $\mathcal{W}^U$ is then formed by removing $\mathcal{D}$ from $\mathcal{S}$ (with $\mathcal{A}$ included as the inserted nodes). This process is illustrated in Figure 2.

*3) Edge Insertion/Deletion*: We generate a random edge stream by sorting all edges of a given dataset into random order. Similar to the node insertion/deletion, we remove the last $k$ edges from the random edge stream to form the original web before any edge modification; we then remove another $l$ edges from the edge stream (edge deletion), and insert the $k$ removed edges back (edge insertion), to form the updated web.

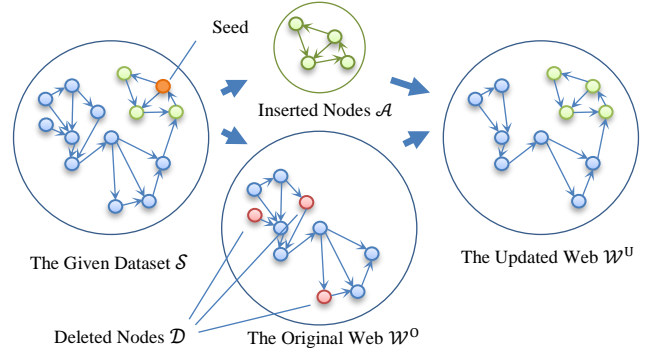

Fig. 2. Dataset generation. Left: the given dataset; Middle: the inserted nodes are fetched from a random seed (orange colored) through BFS, and $\mathcal{W}^O = \mathcal{S}\backslash\mathcal{A}$ forms the original web; Right: some randomly selected nodes from $\mathcal{W}^U$ form the set of deleted nodes $\mathcal{D}$ (red colored, and $\mathcal{W}^U = \mathcal{S}\backslash\mathcal{D}$ makes the updated web ($\mathcal{S}$ with $\mathcal{A}$ inserted and $\mathcal{D}$ removed). Edges are also inserted or deleted following similar procedure.

*4) Parameter Setting*: For each given dataset, we make a node perturbation of 20k nodes being inserted and another 20k nodes being removed. Also, $0.005|E|$ (0.5%) of the total edges are being inserted and another $0.005|E|$ (0.5%) of the edges are being removed at random, which makes approximately 1% edge perturbation in total.

*5) Dangling Nodes*: The dangling nodes are the nodes with no outlinks. Different sulotions have been proposed to handle the dangling nodes in previous works: [17] computes PageRank over only the non-dangling nodes, and the PageRank of the dangling nodes are calculated all at once; [27] excluded the dangling node from the web graph; or like in [21] there's no special treatment for the dangling nodes. In our experiment, we add a link at each dangling node that pointing to the source node $s$. Intuitively, this is equivlent to the random-walk process that we stop at the dangling nodes and reset from the source nodes s, similarly as suggested in [11].

### B. Experiment Environment

All our experiments are conducted on an Ubuntu workstation with an Intel Xeon E5-2680 2.7GHz CPU and 64 GB memory. The algorithms are implemented in C++ 11 using the Eigen library[1], and complied with g++ 5.4 with –O3 option.

### C. Comparison to Existing Methods

We compared our *VWPPR* with the *LazyForwardUpdate* approach [27], a state-of-the-art work that could handle the node insertion/deletion problem to our knowledge. Although in [27] they mainly discussed the PPR computing over undirected graphs, but the *LazyForwardUpdate* is indeed working on directed graphs as described in Algorithm 4 of [27].

For each given dataset, we sample 100 source nodes at random, update PPR for each of the source nodes using both algorithms and calculate the average for comparison. The settings are as below:

*1) Accuracy:* we compute the PPR of the updated network by Gauss-Southwell method with error bound $\epsilon = 10^{-10}$ as the benchmark. The accuracy is defined as the $l_1$-distance of the *VWPPR*/*LazyForwardUpdate* estimations from the benchmark.

---
[1] The Eigen library is retrieved from: http://eigen.tuxfamily.org

*2) Error Bound:* for fairness of the comparison, we adjusted the error bounds of the *LazyForwardUpdate* method and our *VWPPR* method so that their average accuracies are at the same level. For each given dataset, we fix the error bound of the *LazyForwardUpdate* to $\epsilon = 1 \times 10^{-9}$ and compute its accuracy first; and we adjust the error bound of the *VWPPR* method until its accuracy becomes almost the same with the *LazyForwardUpdate*. The detailed settings of the error bounds and the accuracies achieved are listed in Table II.

TABLE II. ERROR BOUND SETTINGS

| Dataset | Method | Error Bound | Avg $l_1$ Error |
|---|---|---|---|
| soc-LiveJournal | VWPPR | $3.3 \times 10^{-9}$ | 0.02933 |
| | LazyFwdUpd | $1 \times 10^{-9}$ | 0.02912 |
| soc-pokec | VWPPR | $6.0 \times 10^{-9}$ | 0.03602 |
| | LazyFwdUpd | $1 \times 10^{-9}$ | 0.03609 |
| enwiki-2018 | VWPPR | $5.4 \times 10^{-9}$ | 0.04334 |
| | LazyFwdUpd | $1 \times 10^{-9}$ | 0.04338 |

TABLE III. COMPARISON OF VWPPR AND LAZYFWDUPDATE

| Dataset | Method | Average Iteration Num | Average Run-time |
|---|---|---|---|
| soc-LiveJournal | VWPPR | $1.9 \times 10^6$ | 2.82s |
| | LazyFwdUpd | $5.17 \times 10^7$ | 8.41s |
| soc-pokec | VWPPR | $2.99 \times 10^6$ | 3.16s |
| | LazyFwdUpd | $3.40 \times 10^7$ | 18.6s |
| enwiki-2018 | VWPPR | $1.28 \times 10^6$ | 3.03s |
| | LazyFwdUpd | $5.55 \times 10^7$ | 10.67s |

*3) Run-time:* we count the time cost of updating the PPR and residual vectors as the run-time. The time cost for organizing the datasets are excluded for both methods.

The comparison result listed in Table III shows that the proposed *VWPPR* method significantly out-performed the *LazyForwardUpdate* method. When both methods achieved almost the same accuracy, for updating PPR under both link and node perturbations, the proposed algorithm is 3 ~ 6 times faster in terms of running time, or 4.4 ~ 10 times faster in terms of iteration numbers. The experiment result shows that the proposed algorithm is very suitable and efficient for PPR updating when node/link insertions and deletions are involved. Under the general case of the network evolvement, especially when node insertion/deletion occurs, we believe updating PPR per every single link change is unnecessary; updating PPR for a small batch of link/node changes could be a better choice.

## V. CONCLUSIONS

In this work, we proposed a novel algorithm named *VWPPR* to handle the node insertion/deletion problem in Personalized PageRank updating. The proposed algorithm is efficient as verified through experiments over real-world datasets, and it also features easy implementation. Instead of updating PPR per single link modification, we believe updating PPR for a small batch of link/node changes is the better choice. The future works include: 1) The virtual web could be integrated with other existing methods, to handle the node modification problem in PPR updating; 2) other optimal virtual web structures may exist that could improve the updating efficiency, which need further study.

REFERENCES

[1] Andersen, R., Chung, F., Lang, K. 2006. Local graph partitioning using pagerank vectors. FOCS:475-86.

[2] Avrachenkov, K., Litvak, N. 2004. Decomposition of the Google PageRank and Optimal Linking Strategy. INRIA.

[3] Avrachenkov, K., Litvak, N. Nemirovsky, D., Osipova, N. 2007. Monte-Carlo methods in pagerank computation: When one iteration is sufficient. SIAM Journal on Numerical Analysis. 45(2):890-904.

[4] Bahmani, B., Chwdhury, A., Goel, A. 2010. Fast Incremental and Personalized PageRank. VLDB;4(3):173-84.

[5] Bahmani, B., Kumar, R., Mahdian, M., Upfal, E. 2012. PageRank on evolving graph. KDD.

[6] Baluja, S., Seth, R., Sivakumar, D., Jing, Y., Yagnik, J., Kumar, S., Ravichandran, D., and Aly, M.. Video suggestion and discovery for youtube: taking random walks through the view graph. In Proceedings of the 17th international conference on World Wide Web, pages 895-904. ACM, 2008.

[7] Berkhin, P. 2006.Bookmark-coloring approach to personalized PageRank computing. Internet Math;3(1):41-62.

[8] Boldi, P., Vigna, S. 2004. The WebGraph Framework: Compression Techniques. ACM Press.

[9] Boldi, P., Rosa, M., Santini, M., Vigna, S. 2011. Layered Label Propagation: A MultiResolution Coordinate-Free Ordering for Compressing Social Networks. ACM Press.

[10] Brin, S., Page, L. The anatomy of a large-scale hypertextual web search engine. 1998. Computer Networks ISDN System.30(1):107-17.

[11] Gleich D. F.. Pagerank beyond the web. SIAM Review, 57(3):321-363, 2015.

[12] Gupta, P, Goel, A., Lin, J, Sharma, A., Wang, D., and Zadeh, R.. Wtf: The who to follow service at twitter. In Proceedings of the 22nd international conference on World Wide Web. International World Wide Web Conferences Steering Committee, 2013.

[13] Haveliwala H. T. Topic-sensitive pagerank: A context-sensitive ranking algorithm for web search. Knowledge and Data Engineering, IEEE Transactions on, 15(4):784-796, 2003.

[14] Internet Live Stats. Available at: http://www.internetlivestats.com/

[15] Jeh G. and Widom J. Scaling personalized web search. In Proceedings of the 12th international conference on World Wide Web. ACM, 2003.

[16] Kamvar, S. D., Haveliwala, T. H., Manning, C. D., Golub, G. H. 2003. Exploting the Block Structure of the Web for Computing PageRank. Technical Report, Stanford University.

[17] Langville, A. N., Meyer, C. D. Deeper Inside PageRank. Internet Mathematics. 2004;1(3):335-80.

[18] Lofgren, P., Banerjee, S., Goel, A., Seshadhri, C. 2014. FAST-PPR: Scaling Personalized PageRank Estimation for Large Graphs. KDD.

[19] Lofgren, P., Banerjee, S., Goel, A. 2015. Bidirectional PageRank Estimation-From Average-Case to Worst-Case. WAW 2016.

[20] Mcsherry, F. 2005. A Uniform Approach to Accelerated PageRank Computation. WWW.

[21] Ohsaka, N., Maehara, T., Kawarabayashi, K.-I. 2015. Efficient PageRank Tracking in Evolving Networks. KDD.:875-84.

[22] Page, L., Brin, S. 1998. The PageRank Citation Ranking: Bringing Order to the Web. Technical Report, Stanford University.

[23] Parreira, J. X., Castillo, C., Donato, D., Michel, S., Weikum, G. The Juxtaposed approximate PageRank method for robust PageRank approximation in a peer-to-peer web search network. The VLDB Journal. 2008;17:291-313.

[24] SNAP Datasets: http://snap.stanford.edu/data

[25] Song, B., Jiang, X., Zhuang, X., PageRank via Virtual Web. Ph.D. Dissertation. University of Missouri-Columbia. 2018.

[26] The WebGraph dataset: http://law.di.unimi.it/datasets.php

[27] Zhang, H., Lofgren, P., Goel, A. Approximate Personalized PageRank on Dynamic Graphs. ACM SIGKDD international conference on Knowledge discovery and data mining; San Francisco, CA, USA2